\documentclass[doublecol,superscriptaddress,floats,floatfix,showpacs]{epl2}

\usepackage{txfonts}
\usepackage{mathrsfs}
\usepackage{amssymb}

\usepackage{graphicx}
\usepackage{dcolumn}
\usepackage{bm}

\def\bra#1{\mathinner{\langle{#1}|}}
\def\ket#1{\mathinner{|{#1}\rangle}}

\def\Bra#1{\left<#1\right|}
\def\Ket#1{\left|#1\right>}

\title{Measuring non-Markovianity of processes with controllable system-environment interaction}

\author{Jian-Shun Tang\inst{1} \and Chuan-Feng Li\inst{1}
$\footnote{email: cfli@ustc.edu.cn}$ \and Yu-Long Li\inst{1}\and Xu-Bo
Zou\inst{1}\and Guang-Can Guo\inst{1}\and Heinz-Peter
Breuer\inst{2}\and Elsi-Mari Laine\inst{3}\and Jyrki Piilo\inst{3}
$\footnote{email: jyrki.piilo@utu.fi}$}

\institute{
  \inst{1} Key Laboratory of Quantum Information, University of Science and Technology of China, CAS, Hefei, 230026, China\\
  \inst{2} Physikalisches Institut, Universit\"at Freiburg, Hermann-Herder-Strasse 3, D-79104 Freiburg, Germany\\
  \inst{3} Turku Centre for Quantum Physics, Department of Physics and Astronomy, University of Turku, FI-20014 Turun yliopisto, Finland
  }
\pacs{03.65.Yz}{Decoherence; open systems; quantum statistical
methods} \pacs{42.50.-p}{Quantum optics} \pacs{03.67.-a}{Quantum
information}

\abstract{Non-Markovian processes have recently become a central
topic in the study of open quantum systems. We realize
experimentally non-Markovian decoherence processes of single photons
by combining time delay and evolution in a polarization-maintaining
optical fiber. The experiment allows the identification of the
process with strongest memory effects as well as the determination
of a recently proposed measure for the degree of quantum
non-Markovianity based on the exchange of information between the
open system and its environment. Our results show that an
experimental quantification of memory in quantum processes is indeed
feasible which could be useful in the development of quantum memory
and communication devices.}

\begin{document}

\maketitle

\section{Introduction}
The theory of open quantum systems describes how a system of
interest is influenced by the interaction with its environment
\cite{Breuer2002}. This interaction often leads to a loss of the
quantum features of physical states and has a great impact on the
dynamical behavior of the open system due to the non-unitary
character of the time evolution. Since any realistic physical system
is coupled to its surroundings, open quantum systems and their
description plays an important role in many applications of modern
quantum physics.

While Markovian, or memoryless, quantum processes are well
understood in the framework of the theory of quantum Markovian
master equations developed during the 70's \cite{GORINI,Lindblad},
non-Markovian processes with memory have recently become of central
importance in the study of open systems
\cite{Stockburger2002,daffer,Piilo2007,breuer2,Lidar2009,Kossakowski2009,EISI,nm-paper,Rivas}.
In general, non-Markovian features can arise, e.g., because of a
strong system-environment interaction, structured reservoirs, due to
initial system-environment correlations, or couplings to a
low-temperature environment or spin bath. Moreover, recent
developments in experimental technology allow reservoir
engineering~\cite{Wineland}, study of quantum
correlations~\cite{Xu2000}, and the development of quantum
simulators for open systems \cite{Blatt}.

On one hand, non-Markovian systems are not yet well-understood even
on the fundamental level, and many efforts have been devoted to the
development of theoretical tools for the treatment of quantum
dynamics with memory
\cite{Piilo2007,breuer2,Lidar2009,Kossakowski2009,EISI,nm-paper,Rivas}.
As a matter of fact, the very definition of quantum non-Markovianity
has been recently under a vivid discussion
\cite{EISI,nm-paper,Rivas}. On the other hand, there are indications
that non-Markovianity may play a role, e.g., in energy transport in
certain photosynthetic complexes \cite{Fleming,Rebentrost} and can
be exploited for quantum metrology \cite{HuelgaM} and cryptography
\cite{NMCrypto}.

In this Letter we implement non-Markovian processes for single
photons and demonstrate that it is possible to determine
experimentally the amount of memory in the system. Our work is based
on a recent theoretical proposal to study the information flow
between the system and its environment, and in particular to
quantify non-Markovianity with the help of the backflow of
information from the environment to the system \cite{nm-paper}. This
allows us to identify a specific process having the largest memory
among all experimentally implemented processes, and then to quantify
the amount of non-Markovianity for this process. There exist earlier
experimental works on dephasing in photonic systems
\cite{Kwiat2000,Berglund2000}, and results which may be regarded as
indications for non-Markovian behavior \cite{Haroche1,Haroche2}.
However, the recent theoretical progress  is opening the path for
rigorous experimental tests of quantum non-Markovianity, for
example, by controlling the initial state of the
environment~\cite{Liu2011}, or by modifying the interaction between
the system and the environment, as is done here. We also note that a theoretical proposal to witness initial system-environment correlations by studying information flow between the system and the environment \cite{Laine2010} has been recently realized experimentally \cite{Li2011, Smirne2011}.

\section{Theoretical framework}
We consider a pure dephasing quantum process for which the density
matrix $\rho$ of the open system evolves according to the master
equation
\begin{equation}
\label{ME-1} \frac{d\rho(t)}{dt} = -i
\frac{\epsilon(t)}{2}\left[\sigma_z,\rho(t)\right]
+\frac{\gamma(t)}{2}\left(\sigma_z\rho(t)\sigma_z - \rho(t)\right).
\end{equation}
Here, $\epsilon(t)$ represents the time-dependent energy shift and
$\gamma(t)$ the time-dependent rate of the decay channel described
by the Pauli operator $\sigma_z$. Generally, non-Markovian dynamics
can be described, e.g., by memory-kernel equations or time local
master equations where the decay rates depend on
time~\cite{Breuer2002}, and we have chosen the latter due to its
conceptual simplicity for the current purpose. Recently, time local
equations have received a great deal of
attention~\cite{Piilo2007,Kossakowski2009,nm-paper,Rivas,Cui2008},
and for a single channel system, as is the case here, one can
directly associate the non-Markovianity with the appearance of
negative periods of the decay rate \cite{nm-paper,Kossakowski2009}.

The open two-state system we consider consists of the horizontal and
the vertical polarization states of a photon, $\ket{H}$  and
$\ket{V}$, respectively. The dephasing process influences the
coherences between the polarization components and the evolution can
be described by the decoherence function $\kappa(t) =
\exp[-\int_0^{t}dt'(\gamma(t')+i \epsilon(t'))]$ which is connected
to the energy shift and the decay rate of the master equation
(\ref{ME-1}) by the relations
\begin{equation} \label{Eq:MEgamma}
 \epsilon(t) = -\Im\left[\dot{\kappa}(t)/\kappa(t)\right], \qquad
 \gamma(t) = -\Re\left[\dot{\kappa}(t)/\kappa(t)\right].
\end{equation}
The corresponding dynamical map $\Phi_t$ which maps the initial
polarization state $\rho(0)$ to the state $\rho(t)=\Phi_t\rho(0)$ at
time $t$ is then given by
\begin{eqnarray} \label{Eq:map}
\rho_{\rm H,H}(t) &=& \rho_{\rm H,H}(0), \quad  \rho_{\rm V,V}(t) = \rho_{\rm V,V}(0), \nonumber \\
\rho_{\rm H,V}(t) &=& \kappa^{\ast} (t) \rho_{\rm H,V}(0), \quad
\rho_{\rm V,H}(t) = \kappa (t) \rho_{\rm V,H}(0). \nonumber
\end{eqnarray}

\begin{figure}[tb]
\includegraphics[scale=0.35]{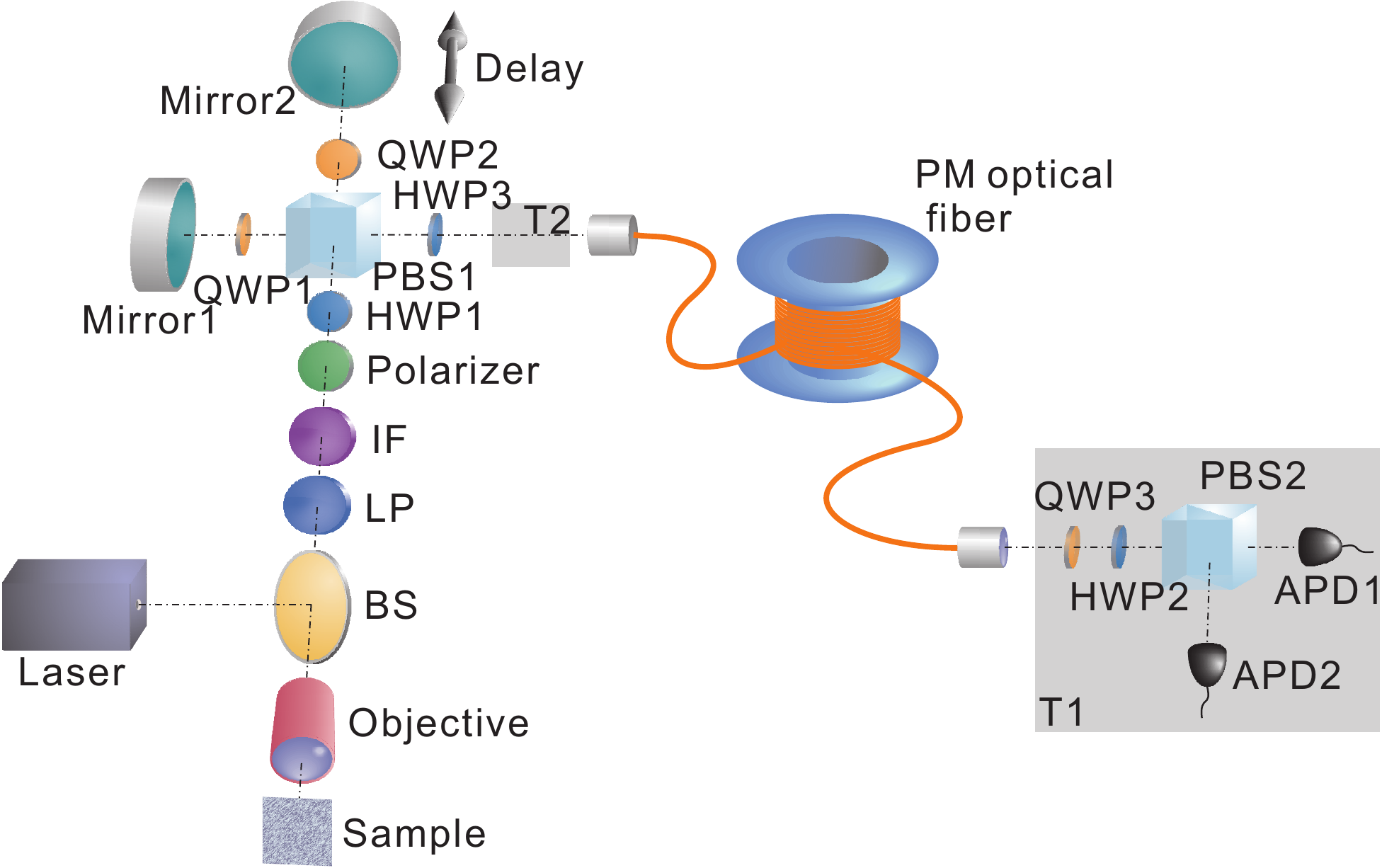}
\caption{\label{Fig1} (Color online) The experimental setup
consisting of three parts: Preparation of a single photon state,
evolution in the delay setup followed by the PM optical fiber, and
state tomography (T1 and T2). Used abbreviations: BS, beamsplitter; LP, long pass filter; IF, interference filter; HWP, half-wave plate; PBS, polarization beam splitter; QWP, quarter-wave plate; PM, polarization maintaining; APD, avalanche photodiode.}
\end{figure}

The  environment consists of the frequency degrees of freedom
$\ket{\omega}$ of the photon and we consider an initial product
state between the system and the environment $\rho_{\rm tot}(0) =
\rho(0) \otimes \int d\omega \int d\omega' A(\omega)
A^{\ast}(\omega') \ket{\omega}\bra{\omega'}.$ The function
$G(\omega)= |A(\omega)|^2$ represents the normalized frequency
distribution of the photon which in our experiments is a Lorentzian
distribution with central frequency $\omega_0$ and full width at
half maximum $2\delta\omega$ (FWHM), $G(\omega) = \frac{\delta
\omega}{\pi}
 \frac{1}{\left( \omega - \omega_0 \right)^2 +\delta \omega^2 }$.
We investigate the case in which the evolution of the frequency
distribution depends on the polarization state such that the total
system dynamics is described by the unitary operator
\begin{eqnarray} \label{Eq:U}
 U_{\rm tot} (t) &=& \Ket{H}\Bra{H} \otimes
 \int d\omega e^{-i\omega u_{\rm H}(t) t} \Ket{\omega}\Bra{\omega}
 \nonumber \\
 &+& \Ket{V}\Bra{V} \otimes
 \int d\omega e ^{-i\omega u_{\rm V}(t) t} \Ket{\omega}\Bra{\omega},
\end{eqnarray}
where the factors $u_{\rm H}(t)$ and $u_{\rm V}(t)$ may depend on
time.

In the experiment we determine the measure for the degree of
non-Markovianity constructed in Ref.~\cite{nm-paper}. For a given
quantum process of an open system described by a dynamical map
$\Phi_t$ this measure is defined by
\begin{equation} \label{MEASURE}
 {\mathcal{N}}(\Phi) = \max_{\rho_{1,2}(0)} \int_{\sigma > 0}
 dt \; \sigma(t,\rho_{1,2}(0)).
\end{equation}
Here, $\rho_{1,2}(0)$ are two initial states of the open system and
$\sigma(t,\rho_{1,2}(0)) = \frac{d}{dt}D(\rho_1(t),\rho_2(t))$ is
the rate of change of the trace distance  $D(\rho_1,\rho_2) =
\frac{1}{2}{\mathrm{tr}}|\rho_1-\rho_2|$. The trace distance
represents a measure for the distinguishability of two quantum
states. In Eq.~(\ref{MEASURE}) the time integral is extended over
all intervals in which the trace distance increases and the maximum
is taken over all pairs of initial states. The measure
${\mathcal{N}}(\Phi)$ thus quantifies the maximal total increase of
the trace distance during the time evolution, which can be
interpreted as the maximal total amount of information that flows
from the environment back to the open system \cite{nm-paper}. In our
experiment the increase of the trace distance signifying
non-Markovian behavior is restricted to a single time interval
$[t_0,t_1]$ which yields $ {\mathcal{N}}(\Phi)=\max_{\rho_{1,2}(0)}
 \big[ D(\rho_1(t_1),\rho_2(t_1))-D(\rho_1(t_0),\rho_2(t_0))\big]$.

\section{Experimental framework}
We measure the non-Markovianity of a quantum evolution process given
by the master equation (\ref{ME-1}) which is implemented for single
photons emitted from a quantum dot (QD). The experimental setup is
shown in Fig.~\ref{Fig1} and includes three parts: Preparation, time
evolution, and state tomography. The first part consists of the
generation of a single photon and its preparation in the pure
initial state
\begin{equation}
\label{Eq:psi} \Ket{\Psi} = \cos(\phi) \ket{H} + \sin(\phi) \ket{V},
\end{equation}
where $\phi$ is the angle of the polarization direction of the
photon from the horizontal direction. A self-assembled InAs/GaAs QD
sample (at a temperature of $7{\mathrm K}$) \cite{Dou2008,Tang2009}
provides the single photon source. A He-Ne laser pumps the sample
through a beam splitter (BS, high transmission efficiency of $92$\%)
and a 50X objective. The photon is collected by the same objective
and separated by a $785$nm long pass filter (LP) and an interference
filter (IF, the central wavelength of the pass band is tunable
around $950$nm and the FWHM is $0.7$nm). Then, a polarizer and a
half-wave plate (HWP1) prepares the photon in the initial state
(\ref{Eq:psi}), where $\phi$ can be changed by rotating HWP1.

The time evolution consists of two main contributions: A delay setup
and a polarization maintaining (PM) optical fiber with a HWP3
between the two contributions. The delay setup includes a
polarization beam splitter (PBS1), two $45^{o}$ (from the horizontal
direction) placed quarter-wave plates (QWP1 and QWP2), and two
mirrors whose relative delay can be controlled. The decoherence
function can generally be written as $\kappa(t) = \int d\omega \,
G(\omega) \exp(i\omega t)$. Considering the delay setup and using
the earlier mentioned Lorentzian photon frequency distribution, we
find
\begin{equation} \label{Eq:kdelay}
\kappa(t) = \exp[i\omega_0 t - \delta\omega |t|],
\end{equation}
where we use the notation $t=2x/c$ with $x$ corresponding to the
delay length between the mirrors. The maximal value of the delay for
a given set of experimental runs is denoted by $x_0$ and the
corresponding time by $t_0 = 2x_0/c$.

In the second phase of the evolution for times $t\geqslant t_0$ the
photon travels in an optical fiber with length $l=100$m and a
designed birefringence of $\Delta n_0=3.5\times10^{-4}$. The total
system evolves according to the Eq.~(\ref{Eq:U}) where now $u_{\rm
H}(t) = n_{\rm H} / \bar{n}$ and $u_{\rm V}(t) = n_{\rm V} /
\bar{n}$, with $\bar{n}$ denoting the average of the ordinary and
the extraordinary refractive indices $ n_{\rm H}$ and $n_{\rm V}$,
respectively. The birefringence of the PM fiber influences the
relative phases of the polarization components
\cite{Kwiat2000,Berglund2000} and by tracing out the environment
leads to the decoherence function
\begin{equation} \label{Eq:kfiber}
 \kappa(t) =
 \exp \left[ i\omega_0 \left(t_0 -
 \frac{\Delta n}{\bar{n}}(t-t_0)\right)
 - \delta \omega  \Big| t_0 - \frac{\Delta n}{\bar{n}}(t-t_0) \Big|\right]
\end{equation}
which is used for times $t_0\leqslant t \leqslant t_{\mathrm f}$,
where $t_{\mathrm f}$ is the termination time of the process when
the photon exits the fiber, and $\Delta n$ is the real
birefringence.

\begin{figure}[tb]
\includegraphics[scale=0.7]{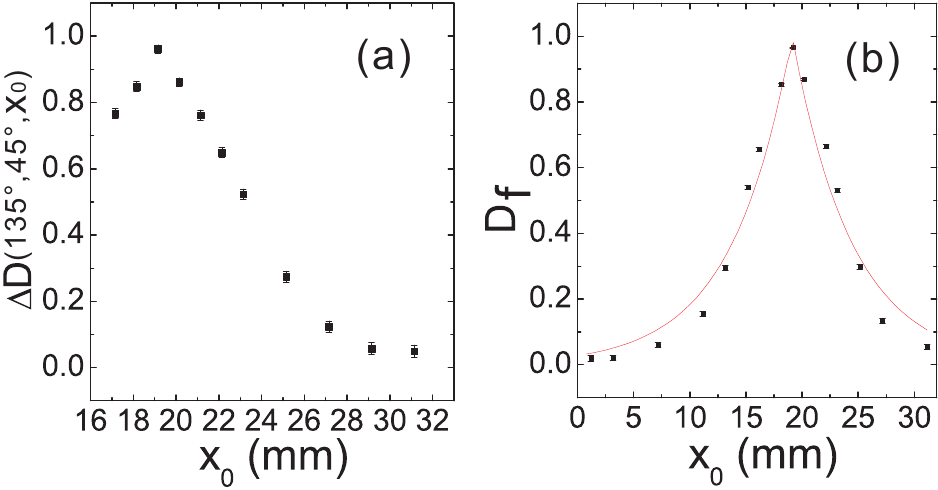}
\caption{\label{Fig2} (Color online) (a) Difference of the trace
distances at times $t_{\mathrm f}$ and $t_0$ for various delay
lengths $x_0$. The initial states are prepared with a fixed pair of
angles $(135^{o},45^{o})$. (b) Final trace distance at time $t_{\rm
f}$ for the pair $(135^{o},45^{o})$. The red solid line is the
theoretical curve (see the text) giving $1/\delta\omega = 35.8\pm
1.9$ps. }
\end{figure}

In the third and last part of the setup a tomography
\cite{James2001} of the final states, i.e., of the states at the end
of the fiber is carried out (T1). We also perform a tomography of
the states after the delay setup (T2). Both tomography setups are
identical utilizing two single-photon avalanche photodiodes (APDs).
The count rate is about $7000$/s, and the integration time is $4$s.

While the length of the fiber is fixed, we can control the maximal
delay $t_0$ before the photons enter the fiber. The evolution of the
reduced system follows the master equation (\ref{ME-1}) for all
times, where $\gamma(t)$ is given by
\begin{eqnarray}
\gamma(t) = \left\{
\begin{array}{ll}
\delta \omega,& 0\leqslant t < t_0, \\
-\delta\omega \frac{\Delta n}{\bar{n}},&  t_0\leqslant t < t_1, \\
\delta\omega \frac{\Delta n}{\bar{n}},&  t_1\leqslant t < t_{\mathrm
f}.
\end{array}
\right.
\end{eqnarray}
These can be obtained by using Eqs.~(\ref{Eq:MEgamma}) and
(\ref{Eq:kdelay}-\ref{Eq:kfiber}), and $t_1= (1+\bar{n}/\Delta n)
t_0>t_0$. We can control the length of the negative decay period by
changing $t_0$.
Note that if $t_0$ is small enough, then $t_{\mathrm f}<t_1$ and the
last positive period does not appear since the decay rate is
negative during all of the fiber evolution.
It is also easy to see from Eq.~(\ref{Eq:kfiber}) that the period of
a negative decay rate corresponds to an increase of $|\kappa(t)|$.
By fixing $t_0$, corresponding to a fixed delay $x_0$, one also
fixes the dynamical map which we denote by $\Phi^{x_0}$, and there
exists a different process for each value of  $x_0$. We are now
interested which one of the processes $\Phi^{x_0}$ has the highest
value of non-Markovianity and what is the corresponding value of
$x_0$.

\section{Results}
To identify the process having the largest value of
non-Markovianity, we first fix the angles of the initial-state pair
as $(\theta,\xi)=(135^{o},45^{o})$ [see Eqs.~(\ref{MEASURE}) and
(\ref{Eq:psi})]. After the delay time $t_0$, the photons travel in
the fiber and a state tomography is performed both before and after
the evolution in the fiber. The difference of the trace distances at
the corresponding times $t_{\rm f}$ and $t_0$ is denoted by $\Delta
D(\theta,\xi,x_0)$ and the result is shown in Fig.~\ref{Fig2}(a). We
find that the maximum is located at $x_0=19.15$mm and takes the
value $\Delta D (135^{o},45^{o},19.15{\rm mm})=0.962\pm0.011$. If we
subtract the dark and background counts of the APDs (about $150$/s),
this value is corrected to $0.988\pm0.011$. The uncertainty is due
to the counting statistics. Fig.~\ref{Fig2}(b) shows the final trace
distance $D_{t_{\mathrm f}}$ as the function of the used delay
$x_0$. By fitting the experimental data to the theoretical curve
$D_{t_{\mathrm f}}(x_0)=\exp(-\delta\omega|\Delta n l - 2x_0|/c)$
[the red solid line in Fig.~\ref{Fig2}(b)] we determine the
parameter $1/\delta\omega = 35.8\pm 1.9$ps.

\begin{figure}[tb]
\includegraphics[scale=0.4]{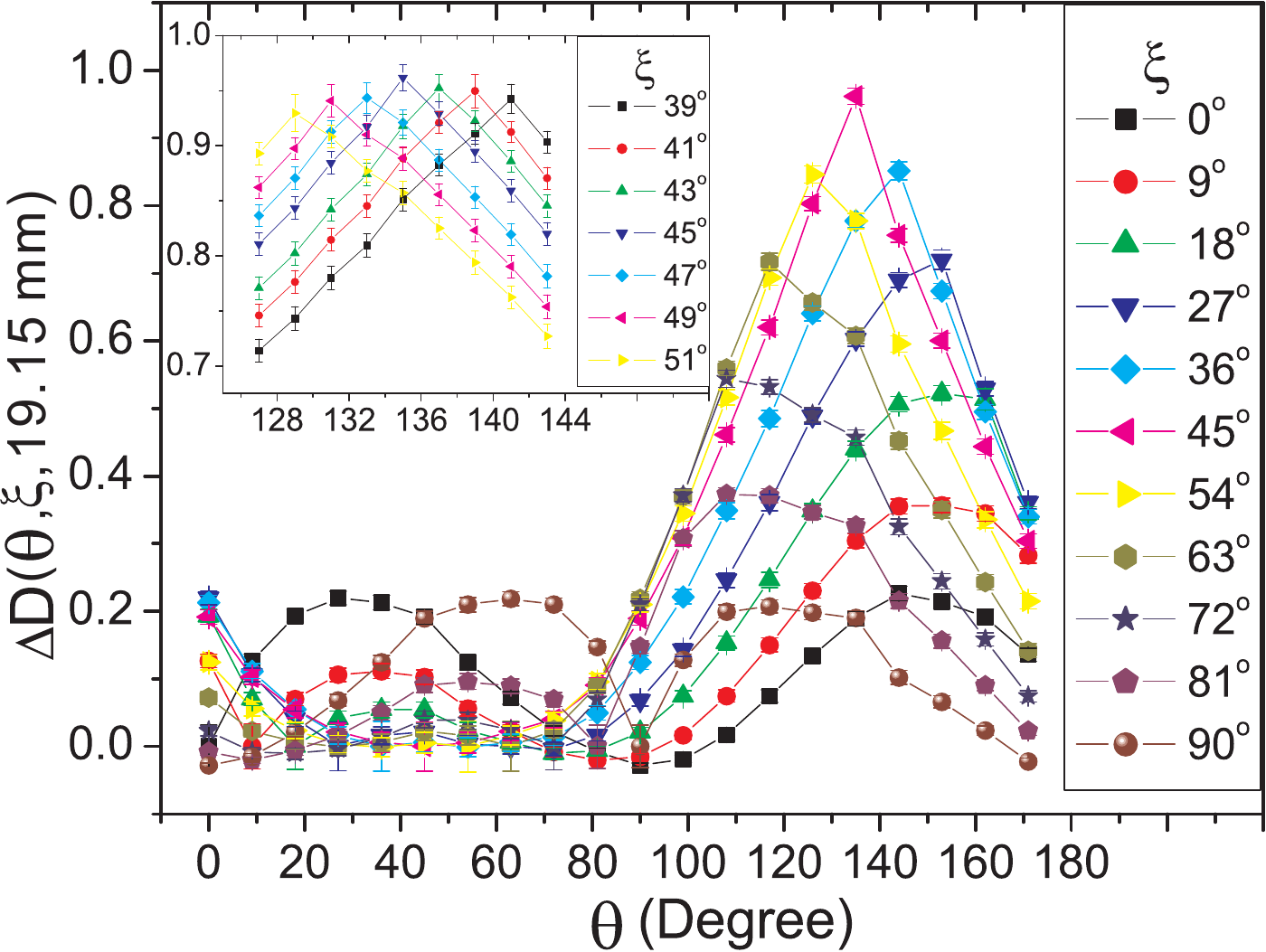}
\caption{\label{Fig3} (Color online) Difference of the final and
initial trace distances. The initial states are prepared with a
fixed delay $x_0=19.15$mm and various pairs of angles
$(\theta,\xi)$. A maximum value is observed for $(\theta,
\xi)=(135^{o},45^{o})$. The inset shows an elaborate measurement
around the pair $(135^{o},45^{o})$. }
\end{figure}

For $x_0\geqslant19.15{\rm mm}$ the decay rate is positive when
$t<t_0$ and negative for all times the photon travels in the fiber,
$t_0\leqslant t \leqslant t_{\rm f}=t_1$. Hence, the results in
Fig.~\ref{Fig2} (a) show that for $x_0\geqslant 19.15{\rm mm}$, the
process $\Phi^{19.15{\rm mm}}$ gives the maximal non-Markovianity
within this group of processes $\Phi^{x_0\geqslant 19.15{\rm mm}}$.
On the other hand, for the processes with $x_0\leqslant 19.15{\rm
mm}$, the smaller $x_0$ the shorter is the negative decoherence
period since the negative decay occurs for $t_0 \leqslant t
\leqslant <t_1 <  t_{\rm f}$. Consequently, the point $x_0=19.15{\rm
mm}$ also gives the maximal non-Markovianity for values
$x_0\leqslant 19.15{\rm mm}$. Thus, we can conclude that
$\Phi^{19.15{\rm mm}}$ yields the maximal non-Markovianity among all
processes experimentally implemented.

In the second step, we fix the delay corresponding to the process
$\Phi^{19.15{\rm mm}}$ and carry out the maximization over pairs of
initial states in the definition (\ref{MEASURE}) of the measure by
changing the pair of angles $(\theta,\xi)$. Experimental results for
$\Delta D (\theta,\xi,19.15{\rm mm})$ are shown in Fig.~\ref{Fig3}.
We clearly see that $\Delta D(135^{o},45^{o},19.15{\rm mm})$ still
represents the maximal value among all pairs of initial states of the form given by Eq.~(\ref{Eq:psi}). We do not perform the maximization over the whole state space, because restricting to pairs given by Eq.~(\ref{Eq:psi}) is sufficient for finding the maximizing pair as we will see later on.

\begin{figure}[tb]
\includegraphics[scale=0.5]{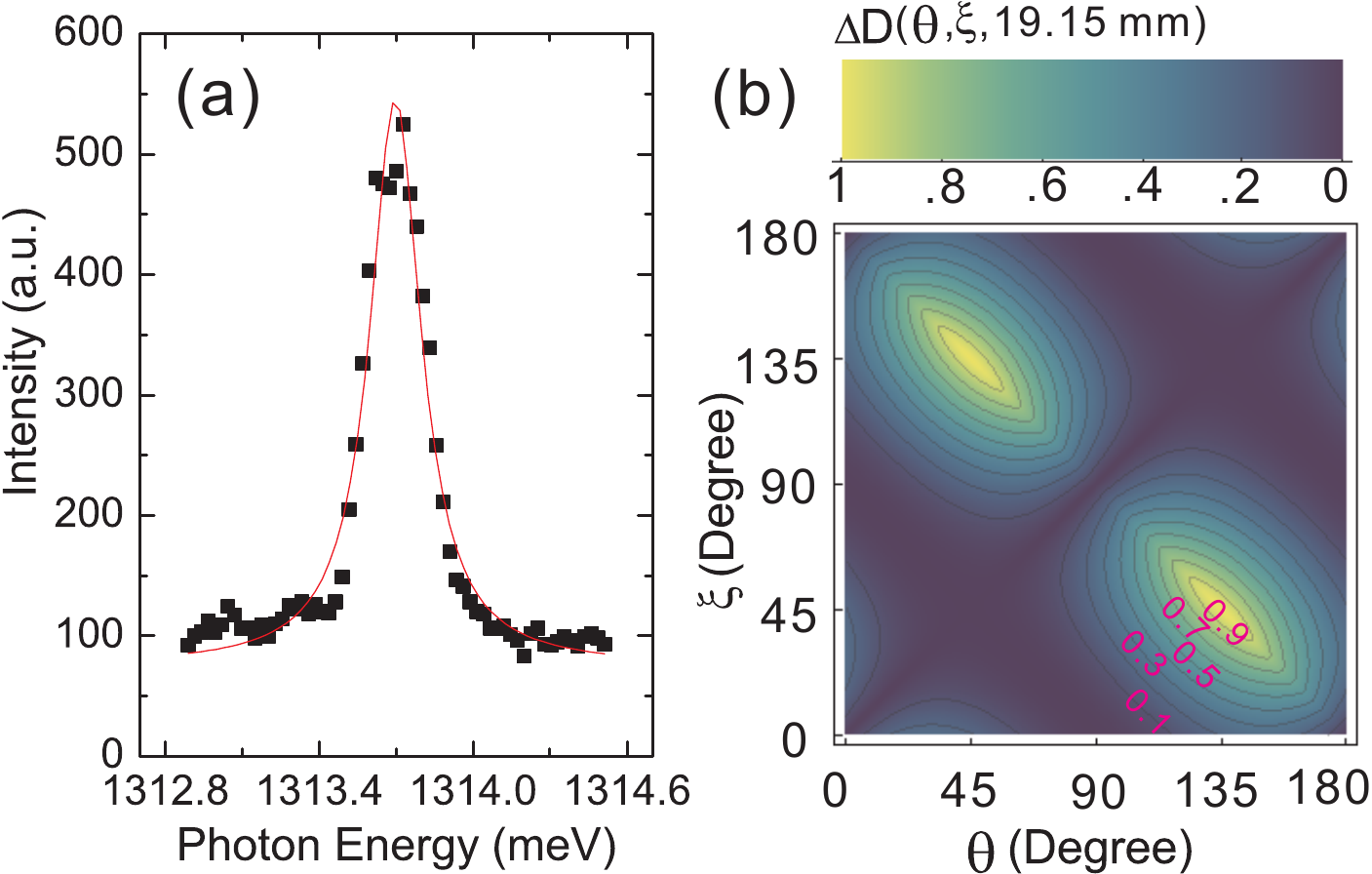}
\caption{\label{Fig4} (Color online) (a) The measured spectrum of
the quantum dot. Fit with Lorentzian curve gives the width
$1/\delta\omega = 34.0\pm 1.7$ps.(b) Simulation results for the
quantity $\Delta D (\theta,\xi,19.15{\rm mm})$ corresponding to the
experimental data of Fig.~\ref{Fig3}. The colors represent the
values of  $\Delta D$. }
\end{figure}

Let us discuss this point in more detail. Firstly, we find that
$2\times 19.15$mm is approximately equal to $\Delta n_0 l = 35$mm.
The discrepancy is explained by noting that the birefringence given
by the manufacturer of the fiber at its designed wavelength $780$nm
is $\Delta n_0=3.5\times10^{-4}$. However, the wavelength of our
chosen peak is at $946.3$nm and consequently the real birefringence
in our experiment is $\Delta n =3.83\times10^{-4}$ [the value used
in Eq.~(\ref{Eq:kfiber})]. This suggests that the birefringence
effect of the PM optical fiber compensates exactly the dephasing
caused by the delay setup when $x_0=19.15$mm.
This means that the open system has recovered all the information
that it lost during the earlier part of the evolution. The
experimentally determined value for the measure of non-Markovinity
is given by $\mathcal{N}(\Phi)=0.962\pm0.012$, and by
$\mathcal{N}(\Phi)=0.988\pm0.011$ when accounting for dark and
background counts. This result matches very closely the theoretical
result $\mathcal{N}(\Phi)=0.972$ obtained by using
Eqs.~(\ref{Eq:kdelay}-\ref{Eq:kfiber}) and the above initial states.
The difference between the experimental and theoretical result may
be caused by mode flips and temperature changes in the PM fiber.
Notice, that increasing the fiber length and correspondingly the delay time ultimately give an increase of 1 for the pair of angles $(135^{o},45^{o})$. Since the trace distance is by definition bounded between 0 and 1 there exists no pair exceeding this value of increase and thus the pair with the angles $(135^{o},45^{o})$ must be maximizing.

We can also directly measure the spectrum of the quantum dot, see
Fig.~\ref{Fig4}(a). The obtained width $1/\delta\omega = 34.0\pm
1.7$ps matches very well the theoretical value $1/\delta\omega =
35.8\pm 1.9$ps [see Fig.~\ref{Fig2}(b)]. Moreover, we have performed
numerical simulations of $\Delta D(\theta,\xi,19.15{\rm mm})$ shown
in Fig.~\ref{Fig4}(b). The found maximum value at
$(135^{\circ},45^{\circ})$ coincides with our experimental result in
Fig.~\ref{Fig3}.

\section{Conclusions}
We have experimentally realized a family of pure decoherence
processes in a non-Markovian open quantum system. The setup allows
to identify the process with strongest memory effects in this family
and to determine the corresponding measure for non-Markovianity
(\ref{MEASURE}). This clearly demonstrates the measurability of this
quantity, including in particular the maximization over the initial
states. Our results show that wide scale experimental studies on
fundamental and practical aspects of quantum systems with memory are
indeed becoming feasible in the wake of the recent vivid theoretical
debate. Moreover, our results may be helpful in the study of quantum
communication processes \cite{Xu}.

\acknowledgments This work was supported by the National Basic
Research Program, National Natural Science Foundation of China
(Grant Nos.~60921091, 10874162, 10734060), the German Academic
Exchange Service, the Graduate School of Modern Optics and
Photonics, and the Magnus Ehrnrooth Foundation.

\end{document}